\documentclass[prd,amsfonts,noshowpacs,nofootinbib]{revtex4}
\usepackage{amsmath}

\begin{document}
\title{Lorentz boost and non-Gaussianity in multi-field
DBI-inflation}
\author{Shuntaro Mizuno\footnote{shuntaro.mizuno@nottingham.ac.uk}$\sharp$}
\author{Frederico Arroja\footnote{arrojaf@yukawa.kyoto-u.ac.jp}$\flat$}
\author{Kazuya Koyama\footnote{Kazuya.Koyama@port.ac.uk}$\natural$}
\author{Takahiro Tanaka\footnote{tanaka@yukawa.kyoto-u.ac.jp}$\flat$}
\affiliation{
$\sharp$School of Physics and Astronomy, University of Nottingham, University Park, Nottingham NG7 2RD, UK;
Research Center for the Early Universe (RESCEU), Graduate School of Science, The University of Tokyo, Tokyo 113-0033, Japan.
\\
$\flat$Yukawa Institute for Theoretical Physics, Kyoto University, Kyoto 606-8502, Japan.
\\
$\natural$Institute of Cosmology and Gravitation, University of Portsmouth, Portsmouth PO1 3FX, UK.
}

\date{\today}

\begin{abstract}
We show that higher-order actions
for cosmological perturbations in the multi-field DBI-inflation
model are obtained by a Lorentz boost
from the rest frame of the brane to
the frame where the brane is moving.
We confirm that this simple method
provides the same third- and fourth- order actions
at leading order in slow-roll
and in the small sound speed limit
as those obtained by the usual ADM formalism.
As an application, we
compute the leading order connected four-point
function of the primordial curvature perturbation
coming from the intrinsic fourth-order contact interaction
in the multi-field DBI-inflation model.
At third order, the interaction Hamiltonian arises purely
by the boost from the second-order action in the rest frame of the brane.
The boost acts on the adiabatic and entropy modes in the same way thus
there exists a symmetry between the adiabatic and entropy modes.
But at fourth order this symmetry
is broken due to the intrinsic
fourth-order action in the rest frame and
the difference between the
Lagrangian and the interaction Hamiltonian.
Therefore,  contrary to
the three-point function, the momentum dependence of the
purely adiabatic component and the components
including the entropic contributions
are different in the four-point function.
This suggests that
the trispectrum can distinguish
the multi-field DBI-inflation model from the single field
DBI-inflation model.
\end{abstract}

\maketitle

\section{Introduction}

Precise measurements of the cosmic microwave background (CMB)
anisotropies such as those obtained by the WMAP satellite \cite{WMAP} provide
valuable information on the very early universe. Any theoretical
model that attempts to explain the evolution of the universe before the
big bang nucleosynthesis will also have to explain the observed CMB
anisotropies. Even though these anisotropies are almost
Gaussian, a small amount of non-Gaussianity is still allowed by the data
\cite{Yadav:2007yy,Komatsu:2008hk,Smith:2009jr}.
The information contained in this non-Gaussian component
will contribute to a huge advance in our understanding of
the very early universe.
For example,
the simplest slow-roll single field inflation models
predict that the non-Gaussianity of
the fluctuations should be very difficult
to be detected \cite{Maldacena:2002vr},
even in future experiments
such as PLANCK \cite{Planck}.
If we detect large non-Gaussianity,
this means that the simplest model of
slow-roll single field inflation would be rejected.

Recently, theoretical models which can produce
sizeable non-Gaussianity has been
extensively studied by many authors
\cite{Linde:1996gt,Bartolo:2001cw,Bernardeau:2002jy,Bernardeau:2002jf,Dvali:2003em,Creminelli:2003iq,Alishahiha:2004eh,Gruzinov:2004jx,Bartolo:2004if,Enqvist:2004ey,Seery:2005wm,Seery:2005gb,Jokinen:2005by,Lyth:2005qk,Salem:2005nd,Seery:2006js,
Sasaki:2006kq,Malik:2006pm,Barnaby:2006cq,Alabidi:2006wa,Chen:2006nt,Huang:2006eh,Chen:2006xjb,Alabidi:2006hg,Seery:2006vu,Byrnes:2006vq,Suyama:2007bg,Arroja:2008ga,Arroja:2008yy,Langlois:2008wt,Langlois:2008qf,Seery:2008ax,
Sasaki:2008uc,Byrnes:2008wi,Byrnes:2008zy,Dutta:2008if,Naruko:2008sq,Suyama:2008nt,Gao:2008dt,Cogollo:2008bi,Rodriguez:2008hy,Ichikawa:2008iq,Byrnes:2008zz,Li:2008fma,Langlois:2008vk,Hikage:2008sk,
Kawasaki:2008sn,Huang:2008zj,Gao:2009gd,Cai:2009hw,Langlois:2009ej,Gao:2009bx,Huang:2009xa,Huang:2009vk,Khoury:2008wj}.
 They basically relax one (or more) of the following
standard conditions: single field, slow-roll,
canonical kinetic term  and standard Bunch-Davies vacuum.
In the present work, we will consider the case of multiple fields with
non-canonical kinetic terms.
For example,
in $K$-inflation models where the kinetic term
of the inflaton field is generic,
the sound speed of the perturbations can be much smaller
than $1$ \cite{ArmendarizPicon:1999rj,Garriga:1999vw},
which leads to large non-Gaussianity.

Among the models, the Dirac-Born-Infeld
(DBI) inflation, motivated by string theory,
can also realize large non-Gaussianity
\cite{Silverstein:2003hf,Alishahiha:2004eh,Chen:2004gc,Chen:2005ad,Chen:2005fe,Chen:2006nt}.
In this model, the inflaton is identified
with the position of a moving D3 brane whose dynamics
is described by the DBI action.
Again, due to the non-trivial form of the kinetic term,
the sound speed can be smaller than $1$ and
the non-Gaussianity becomes large. However, recently
it has been pointed out that DBI-inflation driven by
a mobile D3 brane with large non-Gaussianity might
contradict the current WMAP data.
For current and stringent observational constraints and consequences of DBI-inflation see
\cite{Kecskemeti:2006cg,Lidsey:2006ia,Baumann:2006cd,Bean:2007hc,Lidsey:2007gq,Peiris:2007gz,Kobayashi:2007hm,Lorenz:2007ze,Bean:2007eh,Bird:2009pq}.

One way to avoid this constraint is to consider
multi-field DBI models \cite{Langlois:2008wt}.
Since the position of the brane in each compact direction is
described by a scalar field, DBI-inflation is naturally a
multi-field inflationary model \cite{Easson:2007dh}.
As first pointed out by \cite{Starobinsky:1994mh}, in multi-field
inflation models, the curvature perturbation is modified
on large scales due to the entropy perturbation.
Even though there are some works considering
multi-field inflationary models with kinetic terms
depending on  $X=-G_{IJ} \partial_{\mu} \phi^I
\partial^{\mu} \phi^J /2$, where $\phi^I$ are the scalar fields
$(I=1,2,...)$ and $G_{IJ}$ is the metric in the field space
as in the case of K-inflation
\cite{Huang:2007hh,Langlois:2008mn,Gao:2008dt},
the consistent analysis for the entropy modes in the multi-field
DBI-inflation model has started only very recently
\cite{Langlois:2008wt,Langlois:2008qf,Arroja:2008yy,Langlois:2009ej,Gao:2009gd}.

In \cite{Langlois:2008wt,Langlois:2008qf,Arroja:2008yy,Langlois:2009ej},
the three-point function in the small sound speed limit and
at leading order in the slow-roll expansion was obtained
and it was shown that in addition to the purely adiabatic
three-point function, there exists a mixed component
$\langle
Q_{\sigma}(\mathbf{k_1})Q_{s}(\mathbf{k_2})Q_{s}(\mathbf{k_3})
\rangle$ where $Q_\sigma$ and $Q_s$ are the adiabatic
and the entropy perturbations, respectively.
Since the momentum dependence of the three-point function
from the adiabatic modes was shown to be identical
with the mixed component, the shape of the bispectrum of
the curvature perturbations remains the same as in the single-field
case, while the amplitude is affected by the entropy
perturbation.

The previous works on the non-Gaussianity
in the multi-field DBI-inflation model are limited to the
bispectrum, with the exception of \cite{Gao:2009gd},
where the authors
compute the leading order trispectrum
based on the assumption that it is mainly from
some limited terms of the entropy perturbations.
It is expected that the cosmic microwave background (CMB) trispectrum also provides strong constraints
on early universe models.
At the moment, the constraints are rather weak given by
$|\tau_{NL}| < 10^8$ \cite{Boubekeur:2005fj,Alabidi:2005qi},
where $\tau_{NL}$ denotes the size of the trispectrum.
However, PLANCK will tighten the constraints
significantly reaching
$|\tau_{NL}| \sim 560$ \cite{Kogo:2006kh}.
It is also worth noting that
the analysis in the single field DBI-inflation model
shows that the trispectrum is enhanced
in the small sound speed limit as $\tau_{NL} \sim 1/c_s^4$
\cite{Huang:2006eh,Chen:2009bc,Us}.
As in the bispectrum case, the constraints depend
on the shape of the wave vectors' configuration
\cite{Babich:2004gb}.
Therefore, it is important
to calculate the shape dependence of the trispectrum
in the multi-field DBI-inflation model.
For the details of the observations
of the CMB trispectrum, see
\cite{Hu:2001fa,Okamoto:2002ik,Creminelli:2006gc}.

In this paper,
we calculate the four-point function of the primordial
curvature perturbation coming
from the intrinsic fourth-order contact interaction
in the multi-field DBI-inflation model
and see whether the momentum dependence of the four-point
function is useful to discriminate
the multi-field DBI-inflation model or not.
In order to obtain the third- and fourth-
order actions for cosmological perturbations,
we propose a simple and intuitive method based on
a Lorentz boost, making use of the special property
of the DBI action.
This does not only provide the fourth-order action
easily, but also explains why the momentum dependence
of the three-point function from adiabatic modes is identical
with the one from the mixed component.

The structure of this paper is as follows. In section
\ref{sec:MODEL}, we describe our model and define
perturbations in the flat gauge. We decompose
the perturbations into the adiabatic and entropy
directions. In section \ref{sec:ACTION_Hamiltonian},
after obtaining the fourth-order
action at leading order in slow-roll and in the small sound
speed limit in terms of the decomposed fields
based on the ADM formalism
\cite{Arnowitt:1960es,Maldacena:2002vr,
Seery:2005wm,Seery:2005gb},
we calculate the fourth-order interaction Hamiltonian.
Then, the four-point functions coming
from the intrinsic fourth-order contact interaction are derived in section \ref{sec:FOURPOINTFUNCTION}.
In section \ref{sec:BOOST}, in order to develop understanding
of the results, we derive the same fourth-order action based on a
Lorentz boost which relates the brane-rest frame and the
brane-moving frame.
Section \ref{sec:CONCLUSIONS} is devoted to the conclusion.

\section{\label{sec:MODEL}The model}

We start with the multi-field DBI-inflation model described by the following action
\cite{Leigh:1989jq}
\begin{eqnarray}
&&S=\frac{1}{2}\int
d^4x\sqrt{-g}\left[R+2\tilde{P}(\tilde{X},\phi^I)\right]\,,
\nonumber\\
&&\tilde{P} (\tilde{X},\phi^I) = -\frac{1}{f(\phi^I)}
\left( \sqrt{1-2f(\phi^I) \tilde{X}}-1 \right)-V(\phi^I)\,,
\label{action}
\end{eqnarray}
where we have set $8 \pi G =1$, $R$  is the Ricci scalar,
$\phi^I$ are the scalar fields $(I=1,2,...,N)$,
$f(\phi^I)$ and $V(\phi^I)$ are functions of
the scalar fields determined by string theory
configurations
and $\tilde{X}$ is defined in terms of the determinant
${\cal D} \equiv \mbox{det} (\delta^\mu_\nu + f G_{IJ}
\partial^\mu \phi^I \partial_\nu \phi^J)$ as
$\tilde{X} = (1- {\cal D})/ (2f).$
Here $G_{IJ}$ is the metric in the field space.
We assume that $\tilde{P}$ is a well behaved function
of $\phi^I$ and $\tilde{X}$.
It is also shown that
$\tilde{X}$ is related to the kinetic terms of the scalar
fields as \cite{Langlois:2008wt,Langlois:2008qf}
\begin{eqnarray}
\tilde{X} &=&  G_{IJ} X^{IJ} -2f  X_I ^{\;[I} X_J ^{\;J]}
+ 4f^2 X_I ^{\;[I} X_J ^{\;J} X_K ^{\;K]}
-8f^3  X_I ^{\;[I} X_J ^{\;J} X_K ^{\;K}
X_L ^{\;L]}\,,\\
X^{IJ} &\equiv& -\frac12 g^{\mu\nu} \partial_{\mu} \phi^I
\partial_\nu \phi^J\,,\;\;\;\;
X_{I}^{\;J} = G_{IK} X^{KJ}\,,
\end{eqnarray}
where the brackets denote antisymmetrization.
It is worth noting that even though
$\tilde{X} = X \,(\,= G_{IJ} X^{IJ})$ in the homogeneous
background, this does not hold if we take into account
the inhomogeneous components.

In the background, we are interested in flat, homogeneous and
isotropic Friedman-Robertson-Walker universes described by the
line element
\begin{equation}
ds^2=-dt^2+a^2(t)\delta_{ij}dx^idx^j, \label{FRW}
\end{equation}
where $a(t)$ is the scale factor.
 The Friedman equation and the
continuity equation read
\begin{equation}
3H^2=E_0, \label{EinsteinEq}
\end{equation}
\begin{equation}
\dot{E}_0=-3H\left(E_0+\tilde{P}_0 \right), \label{continuity}
\end{equation}
where the Hubble rate is $H=\dot{a}/a$, dot denotes derivative with
respect to cosmic time $t$,
$E_0$ is the total energy
of the fields which is given by
\begin{equation}
E_0=2X_0^{IJ}\tilde{P}_{0,X^{IJ}}-\tilde{P}_0 ,\label{energy}
\end{equation}
and the subscript $0$ denotes that the quantity is evaluated
in the background.

For the later convenience, we introduce
the following parameter that characterizes
the motion of the brane in this background
and serves as a Lorentz factor
\begin{eqnarray}
\gamma (\phi^I_0, X_0) \equiv
\frac{1}{\sqrt{1-v^2}}\,,
{\hspace{1cm} \rm with}\, {\hspace{0.5cm}}v \equiv
\sqrt{2f_0 X_0},
\label{def_gamma}
\end{eqnarray}
where we have used the fact that
$\tilde{X}_0 = X_0$.

For this model the
speed of propagation of the scalar perturbations
(``speed of sound"), $c_s$, is given by
\begin{equation}
c_s^2 \equiv \left(\frac{\tilde{P}_{,\tilde{X}}}
{\tilde{P}_{,\tilde{X}}+ 2\tilde{X} \tilde{P}_{,\tilde{X}
\tilde{X}}}\right)_0\,.\label{DBIcs}
\end{equation}
It can be shown that $c_s$ is the inverse of  $\gamma$.

Since we are interested in the inflationary background,
we assume the form of $f(\phi^I)$ and $V(\phi^I)$
are chosen so that the inflation is realized at least
for $60$ e-foldings. In order to characterize
this background,
we define the slow-variation parameters, analogues of the
slow-roll parameters, as
\begin{equation}
\epsilon=-\frac{\dot{H}}{H^2}=
\frac{X_0}
{H^2 c_{s}}, \quad
\eta=\frac{\dot{\epsilon}}{\epsilon H}, \quad
\chi=\frac{\dot{c_s}}{c_sH}.
\label{slow_roll_parameters}
\end{equation}
We should note that these slow-variation parameters are more
general than the usual slow-roll parameters and that the smallness
of these parameters does not imply that the field in rolling
slowly.
We assume that the rate of change of the speed of sound is
small (as described by $\chi$) but $c_s$ is otherwise free to change
between zero and one.

We shall consider perturbations on this background.
We decompose the scalar field $\phi^I$
into the background value $\phi_0^I$ and
the perturbation $Q^I$ in the flat gauge as,
\begin{eqnarray}
\phi^I (x,t) = \phi^I_0 (t) + Q^I (x,t)\,.
\end{eqnarray}

Furthermore, as was done in \cite{Gordon:2000hv},
we decompose the perturbations into instantaneous adiabatic and entropy perturbations,
where the adiabatic direction corresponds to
the direction of the background fields' evolution
while the entropy directions are orthogonal to this.
We introduce an orthogonal basis
${e_{n}^I}$, with $n=1,2,...,N$, in the field space
so that the orthonormality conditions are given by \cite{Arroja:2008yy}
\begin{equation}
e_{n}^I e_{m I} =
\frac{1}{c_s}
\delta_{mn}-
\frac{1-c_s^2}{c_s} \delta_{m1} \delta_{n1}\,,
\label{orthonormality_cond}
\end{equation}
where the adiabatic basis is defined as
\begin{equation}
e^I_1 = \sqrt{\frac{c_s}{2X_0}} \dot{\phi}_0 ^I\,.
\label{cond_ad_base}
\end{equation}
Notice that the length of the basis vector 
$e_1^I$ is $c_s$ and that of the other basis vectors
is $1/c_s$.
If we consider the two-field case ($I=1,2$),
the field perturbations are decomposed
into the adiabatic field $Q_\sigma$ and the entropy field
$Q_s$ as
\begin{eqnarray}
Q^I = Q_\sigma e^I_1 + Q_s e^I_2\,.
\label{int_qsigma_qs}
\end{eqnarray}
Hereafter, for simplicity, we will concentrate on
the two-field case although the extension to more fields is straightforward.

\section{\label{sec:ACTION_Hamiltonian}
Fourth-order action and Hamiltonian at leading order}

We first calculate the fourth-order action for $Q_n$, where the subscript $n$ denotes either $\sigma$ or $s$.
Since we are interested in the leading order shape of
the trispectrum, we keep only the leading order
in the slow-roll approximation, where the values of
the slow-variation parameters defined by
Eq.~(\ref{slow_roll_parameters}) are assumed to be small.
We also assume $c_s \ll 1$ because otherwise the
trispectrum is not observable in the future experiments.
Using these approximations and following the ADM formalism
\cite{Arnowitt:1960es,Maldacena:2002vr,Seery:2005wm,Seery:2005gb},
the action up to fourth order can be calculated as
\begin{eqnarray}
&&S^{(main)}_{(2)}= \frac12 \int dt d^3x
\frac{a^3}{c_s^2}
\biggl[\dot{Q}_\sigma^2+
\dot{Q}_s^2-\frac{c_s^2}{a^2}  \bigg(
 \partial_i Q_\sigma \partial^i Q_\sigma
+ \partial_i Q_s \partial^i Q_s\bigg) \biggr]\,,
\label{two_field_lead_dbi_sec}\\
&&S^{(main)}_{(3)}= \frac12 \int dt d^3 x
\frac{a^3}{\sqrt{2X_0 c_s^7}} \biggl[
\dot{Q}_\sigma^3
+
\dot{Q}_\sigma \dot{Q}_s^2
+ \frac{c_s^2}{a^2} \bigg(
\left( \partial_i Q_s \partial^i Q_s
- \partial_i Q_\sigma \partial^i Q_\sigma
\right) \dot{Q}_\sigma
-2
\left(\partial_i Q_{\sigma} \partial^i Q_s \right)
\dot{Q}_s \bigg)
\biggr]\,,\label{two_field_lead_dbi_thir}\\
&&S^{(main)}_{(4)} = \frac{1}{16} \int dx^3 dt
\frac{a^3}{c_s^5 X_0}
\Biggl[
5 \dot{Q}_\sigma^4 +
6 \dot{Q}_\sigma^2 \dot{Q}_s^2
+\dot{Q}_s^4 -\frac{2 c_s^2}{a^2}
\bigg(
3 \dot{Q}_\sigma^2 \partial_i Q_\sigma \partial^i Q_\sigma
- \dot{Q}_\sigma^2 \partial_i Q_s \partial^i Q_s
+4 \dot{Q}_\sigma \dot{Q}_s
\partial_i Q_\sigma \partial^i Q_s
\nonumber\\
&&\qquad\qquad\qquad\qquad\qquad\qquad\qquad\qquad\qquad\qquad\qquad\qquad\quad
+ \dot{Q}_s^2  \partial_i Q_\sigma \partial^i Q_\sigma
+
\dot{Q}_s^2  \partial_i Q_s \partial^i Q_s
\bigg)\nonumber\\
&&\qquad\qquad\qquad\qquad\qquad\qquad
+ \frac{c_s^4}{a^4}  \bigg(
\left(\partial_i Q_\sigma \partial^i Q_\sigma\right)^2
 -2 \left(\partial_i Q_\sigma \partial^i Q_\sigma\right)
\left(\partial_j Q_s \partial^j Q_s\right)
+4 \left(\partial_i Q_\sigma \partial^i Q_s\right)^2
+\left(\partial_i Q_s \partial^i Q_s\right)^2
\bigg)\Biggr]\,.\nonumber\\
\label{two_field_lead_dbi_four}
\end{eqnarray}

In general, the interaction Hamiltonian is not just 
the opposite sign of the interaction part of Lagrangian
as is explained by the following calculation. 
Following \cite{Huang:2006eh}, we define the Lagrangian density
as
\begin{eqnarray}
{\cal{L}} &=&  f^{(0)}_{a} \dot{\alpha}_1^2 +
f^{(0)}_{b} \dot{\alpha}_2^2+j^{(2)}+g^{(0)}_{a} 
\dot{\alpha}_1^3
+ g^{(0)}_{b} \dot{\alpha}_1  \dot{\alpha}_2^2
+ g^{(2)}_{a} \dot{\alpha}_1 + g^{(2)}_{b}  \dot{\alpha}_2 + 
j^{(3)}
\nonumber\\
&&
+h^{(0)}_{a} \dot{\alpha}_1^4 +
h^{(0)}_{b} \dot{\alpha}_1^2  \dot{\alpha}_2^2 +
h^{(0)}_{c} \dot{\alpha}_2^4
+h^{(2)}_{a} \dot{\alpha}_1^2
+ h^{(2)}_{b} \dot{\alpha}_1 \dot{\alpha}_2
+ h^{(2)}_{c} \dot{\alpha}_2^2+ j^{(4)}\,,
\label{two_field_gen_action}
\end{eqnarray}
where $\alpha_m$ with $m=1,2$ denotes $Q_\sigma$ and $Q_s$ respectively. The $f$'s, $j$'s, $g$'s
and $h$'s are all functions of $\alpha_m(t, {\bf x})$, its spatial
derivative $\partial_i \alpha_m$ and time $t$.
They come from 2nd, 3rd and 4th order action, respectively.
For these functions, the superscripts 
$(0)$, $(1)$, $(2)$, $(3)$, $(4)$,
denote the order in the perturbations 
$\alpha_m$ of these functions. 
The subscripts $a$, $b$, $c$, 
just label the functions.

The momentum densities are given by
\begin{eqnarray}
\pi_1 &\equiv&
\frac{\partial {\cal{L}}}{\partial \dot{\alpha}_1}
=2 f^{(0)}_{a} \dot{\alpha}_1 + 3 g^{(0)}_{a} \dot{\alpha}_1^2
+g^{(0)}_{b} \dot{\alpha}_2^2 + g^{(2)}_{a}
+ 4 h^{(0)}_{a} \dot{\alpha}_1^3 
+ 2 h^{(0)}_{b} \dot{\alpha}_1 \dot{\alpha}_2^2 
+ 2 h^{(2)}_{a} \dot{\alpha}_1
+ h^{(2)}_{b} \dot{\alpha}_2\,,
\\
\pi_2  &\equiv&
\frac{\partial {\cal{L}}}{\partial \dot{\alpha}_2}
=2 f^{(0)}_{b} \dot{\alpha}_2 +
2 g^{(0)}_{b} \dot{\alpha}_1 \dot{\alpha}_2 + g^{(2)}_{b}
+2 h^{(0)}_{b} \dot{\alpha}_1^2 \dot{\alpha}_2
+4 h^{(0)}_{c} \dot{\alpha}_2^3 +  h^{(2)}_{b} \dot{\alpha}_1
+2 h^{(2)}_{c} \dot{\alpha}_2\,.
\end{eqnarray}

Making use of these relations, $\dot{\alpha}_m$
are expressed in terms of $\pi_m$ up to third order as
\begin{eqnarray}
\dot{\alpha}_1 &=& \frac{\pi_1}{2 f^{(0)}_{a}}
+ c^{(2)}_{a} \pi_1^2 + c^{(2)}_{b} \pi_2^2 
+ c^{(3)}_{a} \pi_1^3
+ c^{(3)}_{b} \pi_1 \pi_2^2\,,\qquad
\dot{\alpha}_2 = \frac{\pi_2}{2 f^{(0)}_{b}} 
+ d^{(2)}_{a} \pi_1 \pi_2
+ d^{(3)}_{a} \pi_1^2 \pi_2 + d^{(3)}_{b} \pi_2^3\,,
\end{eqnarray}
where the different $c$'s and $d$'s functions are given by
\begin{eqnarray}
c^{(2)}_{a} &=& -\frac{3 g^{(0)}_{a}}
{8 (f^{(0)}_{a} )^3}-
\frac{g^{(2)}_{a}}{2 f^{(0)}_{a} \pi_1^2}\,,\quad
c^{(2)}_{b} = -\frac{g^{(0)}_{b}}
{8 f^{(0)}_{a} (f^{(0)}_{b})^2}\,,\quad
c^{(3)}_{a} = -\frac{1}{2 f^{(0)}_{a}}
\left(\frac{3 g^{(0)}_{a} c^{(2)}_{a}}{f^{(0)}_{a}}
+ \frac{h^{(0)}_{a}}{2 (f^{(0)}_{a})^3}\right)
-\frac{h^{(2)}_{a}}{2 (f^{(0)}_{a})^2 \pi_1^2}\,,\\
c^{(3)}_{b} &=& -\frac{1}{2 f^{(0)}_{a}}
\left(\frac{3 g^{(0)}_{a} c^{(2)}_{b}}{f^{(0)}_{a}}
+ \frac{g^{(0)}_{b} d^{(2)}_{a}}{f^{(0)}_{b}}
+ \frac{h^{(0)}_{b}}{4 f^{(0)}_{a} (f^{(0)}_{b})^2}\right)
-\frac{h^{(2)}_{b}}{4 f^{(0)}_{a} f^{(0)}_{b} \pi_1 \pi_2 }\,,\quad
d^{(2)}_{a} = - \frac{g^{(0)}_{b}}{4 f^{(0)}_{a} 
(f^{(0)}_{b})^2}
-\frac{g^{(2)}_{b}}{2 f^{(0)}_{b} \pi_1 \pi_2}\,,\\
d^{(3)}_{a} &=& - \frac{1}{2 f^{(0)}_{b}}
\left(\frac{g^{(0)}_{b} d^{(2)}_{a}}{f^{(0)}_{a}}
+ \frac{g^{(0)}_{b} c^{(2)}_{a}}{f^{(0)}_{b}}
+\frac{h^{(0)}_{b}}{4 (f^{(0)}_{a})^2 f^{(0)}_{b}}\right) 
-\frac{h^{(2)}_{b}}{4 f^{(0)}_{a} f^{(0)}_{b} \pi_1 \pi_2}
\,,\nonumber\\
d^{(3)}_{b} &=&- \frac{1}{2 f^{(0)}_{b}}
\left(\frac{g^{(0)}_{b} c^{(2)}_{b}}{f^{(0)}_{b}}
+ \frac{h^{(0)}_{c}}{2 (f^{(0)}_{b})^3}\right)
-\frac{h^{(2)}_{c}}{2 (f^{(0)}_{b})^2 \pi_2^2}\,,
\end{eqnarray}
where again for these functions
the superscripts 
$(0)$, $(1)$, $(2)$, $(3)$,
denote the order in the perturbations 
$\alpha_m$ of these functions and  
the subscripts $a$, $b$, 
just label the functions.

The Hamiltonian density is obtained by plugging the previous expressions into
\begin{eqnarray}
{\cal{H}} =  \pi_1 \dot{\alpha}_1
+ \pi_2 \dot{\alpha}_2-{\cal{L}}\,.
\end{eqnarray}

We then separate ${\cal{H}}$ into a kinematic Hamiltonian
density ${\cal{H}}_0$ which is given by
\begin{eqnarray}
{\cal{H}}_0 = \frac{\pi_1^2}{4 f^{(0)}_{a}}+
\frac{\pi_2^2}{4 f^{(0)}_{b}}-j^{(2)}\,,
\end{eqnarray}
and an interaction Hamiltonian density ${\cal{H}}^{int}$.
To use the interaction picture formalism \cite{Weinberg:2005vy},
$\alpha_m$ and $\pi_m$ in this interaction Hamiltonian
should be replaced by their interaction picture counterparts $\alpha_m^I$ and $\pi_m^I$, which satisfy the free equation of
motion determined by ${\cal{H}}_0$. They also satisfy the usual
commutation relations
\begin{eqnarray}
\left[\alpha_m^I(t,{\bf x}),
\pi_n^I (t,{\bf y})\right]
= i  \delta_{mn} \delta^3 \left({\bf x} - {\bf y}\right)\,.
\end{eqnarray}

Expressing $\pi_m^I$ in ${\cal{H}}^{int}$ in terms of
$\dot{\alpha}_m^I$ using
\begin{eqnarray}
&&\dot{\alpha}_1^I = \frac{\partial {\cal{H}}_0 }
{\partial \pi_1^I} = \frac{\pi_1^I}{2 f^{(0)}_{a}}\,,\qquad
\dot{\alpha}_2^I = \frac{\partial {\cal{H}}_0 }
{\partial \pi_2^I} = \frac{\pi_2^I}{2 f^{(0)}_{b}}\,,
\end{eqnarray}
we finally get the third-order and fourth-order
interaction Hamiltonian densities as
(omitting the label ``$I$'' in the variables
in ${\cal{H}}^{int}$ from now on)
\begin{eqnarray}
&&{\cal{H}}^{int}_{(3)} =
-g^{(0)}_{a} \dot{\alpha}_1^3
- g^{(0)}_{b} \dot{\alpha}_1  \dot{\alpha}_2^2
- g^{(2)}_{a} \dot{\alpha}_1 - g^{(2)}_{b}  \dot{\alpha}_2 
- j^{(3)}\,,\\
\label{two_field_gen_hamil_thir}
&&{\cal{H}}^{int}_{(4)} =
\left(\frac{9 (g^{(0)}_{a})^2}{4 f^{(0)}_{a}} -h^{(0)}_{a}
\right)
\dot{\alpha}_1^4
+\left(\frac{3 g^{(0)}_{a} g^{(0)}_{b}}{2 f^{(0)}_{a}}
+\frac{(g^{(0)}_{b})^2}{f^{(0)}_{b}}-h^{(0)}_{b}\right)
\dot{\alpha}_1^2 \dot{\alpha}_2^2
+\left(\frac{(g^{(0)}_{b})^2}{4 f^{(0)}_{a}} 
-h^{(0)}_{c}\right)\dot{\alpha}_2^4\nonumber\\
&&\qquad\quad+
\left(\frac{3 g^{(0)}_{a} g^{(2)}_{a}}{2 f^{(0)}_{a}}
-h^{(2)}_{a}\right) \dot{\alpha}_1^2
+ \left( \frac{g^{(0)}_{b} g^{(2)}_{b}}{f^{(0)}_{b}} 
-h^{(2)}_{b}\right)\dot{\alpha}_1  \dot{\alpha}_2
+\left(\frac{g^{(0)}_{b} g^{(2)}_{a}}{2 f^{(0)}_{a}} 
-h^{(2)}_{c}\right)\dot{\alpha}_2^2
+\left(\frac{(g^{(2)}_{a})^2}{4 f^{(0)}_{a}} +
\frac{(g^{(2)}_{b})^2}{4 f^{(0)}_{b}}-j^{(4)}\right)\,.
\nonumber\\
\label{two_field_gen_hamil_four}
\end{eqnarray}
As in the single-field case \cite{Huang:2006eh},
while the cubic part of  ${\cal{H}}^{int}$
is the opposite sign of the cubic ${\cal{L}}^{int}$,
this is generally not true at fourth order.
As we will see, the extra terms contribute to the leading
order results for $c_s \ll 1$.
By applying Eq.~(\ref{two_field_gen_hamil_four})
to the two-field DBI-inflation model
given by Eqs.~(\ref{two_field_lead_dbi_sec}),
(\ref{two_field_lead_dbi_thir}) and
(\ref{two_field_lead_dbi_four}), we can obtain
the fourth-order interaction Hamiltonian density as
\begin{eqnarray}
{{\cal H}}_{(4)} ^{int} &=& \frac{a^3}{4 X_0 c_s^5} \dot{Q}_\sigma^4
+ \frac{a^3}{4 X_0 c_s^5} \dot{Q}_\sigma^2 \dot{Q}_s^2
+ \frac{a}{4 X_0 c_s^3} \left(\partial_i Q_s \partial^i Q_s
\right) \dot{Q}_\sigma^2
+ \frac{a}{4 X_0 c_s^3} \left(\partial_i Q_s \partial^i Q_s
\right) \dot{Q}_s^2\,.
\label{two_field_lead_dbi_hamil_four}
\end{eqnarray}

\section{\label{sec:FOURPOINTFUNCTION}
The leading order in slow-roll four-point function}

In this section, we derive the connected four-point functions of the adiabatic
and entropy fields coming from the intrinsic fourth-order contact
interactions at leading order in the slow-roll
expansion
and in the small sound speed limit.
The perturbations are promoted to quantum operators as
\begin{equation}
Q_n(\tau,\mathbf{x})=\frac{1}{(2\pi)^3}\int
d^3\mathbf{k}Q_n(\tau,\mathbf{k})
e^{i\mathbf{k}\cdot\mathbf{x}}\,,
\end{equation}
where $\tau$ denotes conformal time and
\begin{equation}
Q_n(\tau,\mathbf{k})=u_n(\tau,\mathbf{k})a_n(\mathbf{k})+
u^*_n(\tau,-\mathbf{k})a^\dag_n(-\mathbf{k})\,,
\end{equation}
and
$a_n(\mathbf{k})$ and $a^\dag_n(-\mathbf{k})$
are the annihilation
and creation operator{\bf s,} respectively, that satisfy the usual
commutation relations:
\begin{eqnarray}
&&\left[a_n(\mathbf{k_1}),a^\dag_m(\mathbf{k_2})\right]
=(2\pi)^3\delta^{(3)}(\mathbf{k_1}-\mathbf{k_2})\delta_{nm}\,,\quad
\left[a_n(\mathbf{k_1}),a_m(\mathbf{k_2})\right]
=\left[a_n^\dag(\mathbf{k_1}),a_m^\dag(\mathbf{k_2})\right]=0\,.
\end{eqnarray}
At leading order, 
the solutions for the mode functions are given by
\begin{equation}
u_n(\tau,\mathbf{k})
=A_n\frac{1}{k^{3/2}}\left(1+ikc_s\tau\right)e^{-ikc_s\tau}\,.
\end{equation}
The two-point correlation functions are then obtained as
\begin{eqnarray}
&&\langle
0|Q_n(\tau=0,\mathbf{k_1})Q_m(\tau=0,\mathbf{k_2})|0\rangle
=(2\pi)^3\delta^{(3)}(\mathbf{k_1}+\mathbf{k_2})
\mathcal{P}_{Q_n}\frac{2\pi^2}{k_1^3}\delta_{nm}\,,
\end{eqnarray}
where the power spectra $\mathcal{P}_{Q_n}$ are defined as
\begin{equation}
\mathcal{P}_{Q_n}=\frac{|A_n|^2}{2\pi^2}\,, \quad
|A_\sigma|^2=|A_s|^2=\frac{H^2}{2c_{s}}\,,
\end{equation}
and they should be evaluated at the time of the sound horizon crossing
${c_s}_* k_1=a_*H_*$.

In terms of these quantum operators,
the connected four-point correlation function coming from
the contact interaction
in the interaction picture formalism is given by
\cite{Maldacena:2002vr,Weinberg:2005vy}
\begin{eqnarray}
&&\langle\Omega|Q_m(t,\mathbf{k_1})Q_n(t,\mathbf{k_2})
Q_p(t,\mathbf{k_3}) Q_q(t,\mathbf{k_4})|\Omega\rangle
=-i\int^t_{t_0}d\tilde t \langle 0
|\left[Q_m(t,\mathbf{k_1})Q_n(t,\mathbf{k_2})
Q_p(t,\mathbf{k_3}) Q_q(t,\mathbf{k_4}),H^{int}(\tilde
t)\right]|0\rangle\,, \nonumber\\
\end{eqnarray}
where $t_0$ is some early time during inflation
when the fields' vacuum fluctuations are
deep inside the horizon and
$t$ is some time after the horizon exit.
$|\Omega\rangle$ is the interacting vacuum
which is different from the free theory vacuum $|0\rangle$.
If one uses conformal time, it is a good approximation
to perform the integration from $-\infty$ to $0$
because $\tau\approx-(aH)^{-1}$. The interaction Hamiltonian
$H^{int}$ is given by
$H^{int} = \int d^3 x {{\cal H}}^{int}$.

The purely adiabatic, purely entropic and
mixed components
are given by
\begin{eqnarray}
&&
\label{4point_pureadiabatic}
\langle\Omega|Q_\sigma(0,\mathbf{k_1})Q_\sigma(0,\mathbf{k_2})
Q_\sigma(0,\mathbf{k_3})Q_\sigma(0,\mathbf{k_4})
|\Omega\rangle=
(2 \pi)^3 \delta^{(3)} (\sum_{i=1}^4 \mathbf{k_i})
\frac{H^8}{2X_0 c_s^6}
\frac{1}{\Pi_{i=1}^4 k_i^3}
\left( -36 A_1 \right)\,,\\
&& \qquad\qquad\qquad\qquad A_1 = \frac{\Pi_{i=1}^4 k_i^2}{K^5}\,,
\quad K=\sum_{i=1} ^{4} k_i\,,\\
&&
\label{4point_pureentropy}
\langle\Omega|Q_s(0,\mathbf{k_1})Q_s(0,\mathbf{k_2})
Q_s(0,\mathbf{k_3})Q_s(0,\mathbf{k_4})
|\Omega\rangle=
(2 \pi)^3 \delta^{(3)} (\sum_{i=1}^4 \mathbf{k_i})
\frac{H^8}{2X_0 c_s^6}
\frac{1}{\Pi_{i=1}^4 k_i^3}
\biggl(-
\frac{1}{8} A_2\biggr)\,,\\
&&\qquad\qquad\qquad\qquad
A_2 = \frac{k_1^2 k_2^2 (\mathbf{k_3}\cdot\mathbf{k_4})}
{K^3} \left( 1+ \frac{3(k_3 + k_4)}{K}
+ \frac{12 k_3 k_4}{K^2}\right)+{\rm perm.}\label{def_a2}\\
&&
\label{4point_mixed}
\langle\Omega|Q_\sigma(0,\mathbf{k_1})Q_\sigma(0,\mathbf{k_2})
Q_s(0,\mathbf{k_3})Q_s(0,\mathbf{k_4})
|\Omega\rangle=
(2 \pi)^3 \delta^{(3)} (\sum_{i=1}^4 \mathbf{k_i})
\frac{H^8}{2X_0 c_s^6}
\frac{-1}{\Pi_{i=1}^4 k_i^3}
\biggl( 6 A_1 + \frac{1}{2} A_3 \biggr)\,,\\
&&
\qquad\qquad\qquad\qquad
A_3 = \frac{k_1^2 k_2^2 (\mathbf{k_3}\cdot\mathbf{k_4})}
{K^3} \left( 1+ \frac{3(k_3 + k_4)}{K}
+ \frac{12 k_3 k_4}{K^2}\right)\,,
\end{eqnarray}
where ``perm.'' in Eq.~(\ref{def_a2}) denotes
the 23 permutations of the four-momenta.
The purely adiabatic component agrees with the result
of the single-field DBI-inflation model \cite{Huang:2006eh}.

We need to relate the four-point functions of the scalar
fields to the four-point function of the comoving curvature
perturbation $\mathcal{R}$ which is closely related to the observable
quantity.
As in \cite{Arroja:2008yy}, $\mathcal{R}$ and $Q_\sigma$
are related as
\begin{eqnarray}
\mathcal{R} = \frac{\sqrt{c_s} H}{ \sqrt{2X_0}} Q_\sigma\,,
\end{eqnarray}
and it is convenient to define the entropy perturbation
$\mathcal{S}$ as
\begin{eqnarray}
\mathcal{S} = \frac{\sqrt{c_s} H}{ \sqrt{2X_0}} Q_s\,,
\end{eqnarray}
so that the power spectra are
$\mathcal{P}_{\mathcal{S}_*}\simeq\mathcal{P}_{\mathcal{R}_*}$,
where the subscript $*$ means that the quantity should be
evaluated at the sound horizon crossing.

In this work, we ignore the possibility that the entropy
perturbation during inflation can lead to a primordial entropy
perturbation that could be observable in the CMB. But we shall
consider the effect of the entropy perturbation on the final
curvature perturbation. Following the analysis of
\cite{Wands:2002bn}, we describe the conversion of the entropy
perturbation into the curvature perturbation by a transfer
coefficient $T_{\mathcal{RS}}$. Then the final curvature perturbation
is expressed in terms of the adiabatic and entropy
field perturbations as
\begin{eqnarray}
&&\mathcal{R}=\mathcal{A}_\sigma
Q_{\sigma *}+\mathcal{A}_sQ_{s*}\,,
\quad
\mathcal{A}_\sigma=
\left( \frac{\sqrt{c_s} H}{ \sqrt{2X_0}}\right)_*\,,
\quad \mathcal{A}_s= T_{\mathcal{RS}}
\left( \frac{\sqrt{c_s} H}{ \sqrt{2X_0}}\right)_*\,.
\end{eqnarray}

Hence the connected four-point function of $\mathcal{R}$ at
leading order is given by
\begin{eqnarray}
\langle\mathcal{R}(\mathbf{k_1})\mathcal{R}(\mathbf{k_2})
\mathcal{R}(\mathbf{k_3}) \mathcal{R}(\mathbf{k_4})\rangle
 &=& \mathcal{A}^4_\sigma
\langle
Q_\sigma(\mathbf{k_1})Q_\sigma(\mathbf{k_2})
Q_\sigma(\mathbf{k_3}) Q_\sigma(\mathbf{k_4})\rangle
+ \mathcal{A}^2_\sigma \mathcal{A}^2_s
\big(\langle
Q_\sigma(\mathbf{k_1})Q_\sigma(\mathbf{k_2})
Q_s (\mathbf{k_3}) Q_s (\mathbf{k_4})\rangle + {\rm perm.}\big)
\nonumber\\
&& + \mathcal{A}^4_s
\langle
Q_s (\mathbf{k_1})Q_s (\mathbf{k_2})
Q_s (\mathbf{k_3}) Q_s (\mathbf{k_4})\rangle\,,
\label{4point_curv}
\end{eqnarray}
where ``perm'' denotes five permutations of
the four-momenta. This constitutes one of the main results of this work.

Since the mixed component and the purely entropic component
have different momentum dependence from the purely adiabatic
component, the momentum dependence of the
resultant four-point function of the curvature perturbation
is different from that in the single-field
DBI-inflation model \cite{Huang:2006eh}.
The only effect for the bispectrum due to the presence of the multiple fields is a change in its amplitude with respect to the single field case, however for the trispectrum the presence of the multiple fields affects also the shape dependence. So in principle, the trispectrum can be used to distinguish between the multi-field and the single-field DBI-inflation models.

\section{\label{sec:BOOST} Lorentz boost}

In this section, we will present an alternative and simpler method to obtain the leading order\footnote{In fact, with this method one can also obtain the sub-leading terms in $c_s^2$.} action for the perturbations.

It is well known that the DBI action (\ref{action}) describes the
motion of a brane in a higher dimensional spacetime. For simplicity,
let us take $g_{\mu\nu}$ as the Minkowski metric\footnote{In our case
$g_{\mu\nu}$ will be an inflating FRW metric but this is conformally
Minkowski and the final result will be the same as for the Minkowski
case, up to powers of the scale factor.}
and $f$ is constant. Then in the frame where the
brane is at rest in the background (brane-rest frame),
$\tilde X$ will be a small quantity because it is written in terms of perturbations of the brane positions. Then we can expand the Lagrangian (\ref{action}),
by ignoring the potential terms which do not contribute at the leading order as
\begin{equation}
\tilde P=X-\frac{f}{2}X^2+fX_I^JX_J^I+\mathcal{O}(X_{IJ}^3), \label{ExpandedLagrangian}
\end{equation}
where $X$ and $X_{IJ}$ are written in the coordinates of this new frame
that we will denote by
$(\tilde t, \mathbf{x}, \tilde{\sigma},s)$.
$\tilde{\sigma}$ is the
coordinate along the direction of the motion of the brane in the
background (adiabatic direction) and $s$ parameterizes the orthogonal
direction (entropy direction).
In the brane-rest frame,
the DBI action for the two-field model can be
written in a fairly simple form as
\begin{eqnarray}
S_{rest} &=&\int\bigg[-\frac{1}{2}\partial_\mu
\tilde{\sigma}\partial^\mu\tilde{\sigma}-\frac{1}{2}\partial_\mu s\partial^\mu s
+ \frac{f}{8}\bigg(\left(\partial_\mu\tilde{\sigma}\partial^\mu\tilde{\sigma}\right)^2+\left(\partial_\mu s\partial^\mu s\right)^2
+ 4\left(\partial_\mu s\partial^\mu \tilde{\sigma}\right)^2
-2\partial_\mu\tilde{\sigma}\partial^\mu\tilde{\sigma}\partial_\nu s\partial^\nu s\bigg)\bigg]d\tilde td\mathbf{x},\label{ExpandedLagrangian2}
\end{eqnarray}
where we have ignored higher-order terms.

We are interested in the behaviour of the perturbations
in the frame corresponding
to the set-up shown in the previous sections and
for this purpose, it is necessary to know the Lagrangian
of the perturbations in such a frame.
From Eq.~(\ref{def_gamma}),
this frame is the one where the brane is moving
with the velocity $v=\sqrt{2f X_0}$ in the background
(brane-moving frame).
If we label the coordinates of the brane-moving frame
as $(t, \mathbf{x}, \sigma,s)$, the coordinate variables
in these two frames are related by a Lorentz transformation
\begin{equation}
t=\gamma\left(\tilde t+v\sqrt{f}\tilde{\sigma} \right),\quad
\sqrt{f}\sigma= \gamma\left(\sqrt{f}\tilde{\sigma}+
v\tilde t\right),
\label{lorentz_tr}
\end{equation}
and its inverse transformation
\begin{equation}
\tilde t=\gamma\left(t-v\sqrt{f} \sigma \right),\quad
\sqrt{f} \tilde{\sigma}=
\gamma\left( \sqrt{f} \sigma-vt\right),
\label{lorentz_tr_inv}
\end{equation}
with  ${\bf x}$ and $s$ unchanged.
From Eq.~(\ref{lorentz_tr}), in terms of the
background value $\sigma_0$, $v$ can be expressed
as $v^2= f \dot{\sigma}_0^2$, which means
$\sigma_0 = \sqrt{2X_0}$.
We are interested in the small sound speed limit or equivalently when the brane is relativistic, i.e. $v\sim 1$.

In order to see the behaviour of the perturbations
in the brane-moving frame, it is convenient to introduce
a new variable as
\begin{equation}
\delta \sigma(t,\mathbf{x})
\equiv
\sigma(t,\mathbf{x})-\frac{vt}{\sqrt{f}}
=
\frac{\tilde{\sigma}(\tilde t,\mathbf{x})}{\gamma},
\end{equation}
which is nothing but the adiabatic perturbation
in the brane-moving frame, since $vt/\sqrt{f}$
can be interpreted as $\sigma_0$ in the case
$\dot{\sigma}_0$ is constant.
For the entropy perturbation in the brane-moving frame,
we can continue to use $s$, since $s$ is invariant
under the Lorentz transformation.
It is worth noting that these $\delta \sigma$ and $s$
are related with $Q_\sigma$ and $Q_s$ introduced
in Eq.~(\ref{int_qsigma_qs}) as
\begin{equation}
\delta \sigma = \sqrt{c_s} Q_\sigma,\;\;\;\;\;\;\;\;
s = \frac{1}{\sqrt{c_s}} Q_s,
\label{rel_q_sig_sig_q_s_s}
\end{equation}
since $Q_\sigma$ and $Q_s$ are
defined in terms of the basis satisfying
(\ref{orthonormality_cond}) and
(\ref{cond_ad_base}).

For the Lagrangian in the brane-moving frame
it is necessary to know the transformation law of
not only the coordinate variables, but also its derivatives.
Since the time derivative of a quantity $q$
in the brane-moving frame is
\begin{eqnarray}
\frac{\partial q}{\partial t}(t,\mathbf{x})
&=&\frac{\partial q}{\partial \tilde t}(\tilde
t,\mathbf{x})\frac{\partial \tilde t}{\partial t}
=\frac{\partial q}{\partial \tilde t}
(\tilde t,\mathbf{x})\left(\frac{1}{\gamma}
-v\sqrt{f}\gamma \delta\dot{\sigma}\right)
=\tilde\gamma^{-1}\frac{\partial q}{\partial \tilde t}(\tilde t,\mathbf{x}),
\end{eqnarray}
where we define $\tilde \gamma$ as
\begin{eqnarray}
\tilde\gamma&=&
\left(\frac{1}{\gamma}-v\sqrt{f}\gamma \delta\dot{\sigma}
\right)^{-1}
\approx
\gamma\left(1+v\sqrt{f}\gamma^2 \delta\dot{\sigma}
+v^2f\gamma^4( \delta\dot{\sigma})^2\right)
+\mathcal{O}\biggl(( \delta\dot{\sigma})^3\biggr),
\end{eqnarray}
we can express the time derivative of a quantity $q$
in the brane-rest frame by
the one in the brane-moving frame as
\begin{equation}
\frac{\partial q}{\partial \tilde t}(\tilde t,\mathbf{x})=\tilde\gamma\frac{\partial q}{\partial t}(t,\mathbf{x}).
\label{boost_time_deriv}
\end{equation}
Similarly, the spatial gradient of a quantity $q$
in the brane-rest frame can be related with the one
in the brane-moving frame as
\begin{equation}
\nabla q(\tilde t,\mathbf{x})
=\nabla q(t,\mathbf{x})+v\sqrt{f}\gamma\tilde\gamma
\nabla (\delta \sigma)\dot q.
\label{boost_spatial_grad}
\end{equation}

One can now use Eqs.~(\ref{ExpandedLagrangian2}), (\ref{lorentz_tr_inv}),
(\ref{rel_q_sig_sig_q_s_s}), (\ref{boost_time_deriv})
and (\ref{boost_spatial_grad}) to obtain the actions up
to fourth-order in the brane-moving frame as
\begin{equation}
S_{mov}^{(2)}\sim\int\frac{a^3}{2 c_s^2}
\left[\dot{Q}_\sigma^2 + \dot{Q}_s^2-
\frac{c_s^2}{a^2}\left(\partial_i Q_\sigma \partial^i Q_\sigma
+ \partial_i Q_s \partial^i Q_s \right)
\right]dtd\mathbf{x},
\label{2ndorderaction_boost}
\end{equation}
\begin{eqnarray}
S_{mov}^{(3)}&\sim&\int
\frac{a^3}{2 \sqrt{2 X_0 c_s^7}}
\bigg[\dot{Q}_\sigma ^3 + \dot{Q}_\sigma \dot{Q}_s^2-
\frac{c_s^2}{a^2} \left\{ \dot{Q}_\sigma
\left(\partial_i Q_\sigma \partial^i Q_\sigma
- \partial_i Q_s \partial^i Q_s\right)
+ 2 \dot{Q}_s \partial_i Q_\sigma \partial^i Q_s\right\}
\bigg]dtd\mathbf{x},\label{3rdorderaction_boost}
\end{eqnarray}
\begin{eqnarray}
S_{mov}^{(4)}&\sim&\int
\frac{a^3}{16 c_s^5 X_0}\bigg[5 \dot{Q}_\sigma ^4 +
6 \dot{Q}_\sigma ^2 \dot{Q}_s^2 + \dot{Q}_s^4\nonumber\\
&&-\frac{2 c_s^2}{a^2}
\left\{3 \dot{Q}_\sigma^2
\partial_i Q_\sigma \partial^i Q_\sigma
-\dot{Q}_\sigma^2 \partial_i Q_s \partial^i Q_s
+4 \dot{Q}_\sigma \dot{Q}_s
\partial_i Q_\sigma \partial^i Q_s
+ \dot{Q}_s ^2\partial_i Q_\sigma \partial^i Q_\sigma
+ \dot{Q}_s ^2\partial_i Q_s \partial^i Q_s  \right\}
\nonumber\\
&&+\frac{c_s^4}{a^4}
\left\{\left(\partial_i Q_\sigma \partial^i Q_\sigma \right)^2
-2\left(\partial_i Q_\sigma \partial^i Q_\sigma \right)
\left(\partial_i Q_s \partial^i Q_s \right)
+4\left(\partial_i Q_\sigma \partial^i Q_s \right)^2
+ \left(\partial_i Q_s \partial^i Q_s \right)^2\right\}
\bigg]dtd\mathbf{x},
\label{4thorderaction_boost}
\end{eqnarray}
where we introduced the scale factor dependence, we used $d\tilde t=dt/\tilde\gamma$, $v \sim 1$ and
$f\sim1/\dot{\sigma}_0 ^2 \sim 1/(2X_0)$.
Although cubic interactions are absent in the
brane-rest frame (see Eq. (\ref{ExpandedLagrangian2})),
they are induced by the
boost in the brane-moving frame
(see Eq.(\ref{3rdorderaction_boost})).
At fourth order, quartic interactions
are already present in the brane-rest frame but additional terms are
induced in the brane-moving frame.
It is worth noting that
Eqs.~(\ref{2ndorderaction_boost}),
(\ref{3rdorderaction_boost}) and (\ref{4thorderaction_boost})
agree with Eqs.~(\ref{two_field_lead_dbi_sec}),
(\ref{two_field_lead_dbi_thir})
and (\ref{two_field_lead_dbi_four})
derived by the usual
ADM formalism.

Now we are in the position to discuss the momentum dependence of the
bispectrum and trispectrum. The second-order action in the rest frame
of the brane is symmetric under the exchange between the adiabatic and entropy modes.
At quadratic order, 
because of the change of the time coordinate,
$dt\approx \gamma d\tilde t$, the sound speed in the brane-moving frame
deviates from the
speed of light. Since the shift of the sound speed is solely due to this
coordinate change, both the adiabatic and entropy
modes should have a common sound speed~\cite{Langlois:2008qf}.
The third-order action in the brane-moving frame,
which is generated by the boost transformation,
also originates from the second-order action in the brane rest frame.
As is seen from Eqs.~(57) and (58), the boost acts on both modes in the same way.
Therefore the boost transformation preserves the symmetry between the adiabatic
and entropy modes in the sense that the mixed component which contains
$Q_s$ reduces to the purely adiabatic component if all $Q_s$ are replaced
with $Q_\sigma$. At third order, the
interaction Hamiltonian is just
the minus of the third-order interaction Lagrangian.
Thus, at this order, the three-point function from
the mixed component has the same momentum
dependence as in the purely adiabatic case.
Thus, adding the contributions from the entropy modes
does not change the momentum dependence.

At fourth order, there are two contributions
to the action in the brane-moving frame.
One is the fourth-order action
that arises by the boost from the
second-order action in the brane-rest frame:
\begin{equation}
S_{boost}^{(4)} \sim\int
\frac{a^3}{4 c_s^5 X_0}\bigg[
\dot{Q}_\sigma^4+ \dot{Q}_\sigma^2\dot{Q}_s^2
-\frac{c_s^2}{a^2}
\bigg( \dot{Q}_\sigma^2
\partial_i Q_\sigma \partial^i Q_\sigma +
\dot{Q}_s^2
\partial_i Q_\sigma \partial^i Q_\sigma
\bigg)\bigg]dtd\mathbf{x}.
\label{4thfrom2nd}
\end{equation}
As in the case of the third-order action, the boost
preserves the symmetry between the adiabatic and entropy
modes. Thus, the mixed component becomes the same
as the pure adiabatic component by replacing $Q_s$
with $Q_\sigma$ and the mixed component would
give the same shape dependence of the trispectrum as the pure
adiabatic component. However, at fourth order,
there exists the intrinsic
fourth-order action
in the brane-rest frame which contains pure
adiabatic, pure entropy and mixed components.
If one adds these
contributions to Eq.~(\ref{4thfrom2nd}), the symmetry
between the adiabatic mode and entropy mode is broken.
Moreover, at fourth order, the interaction
Hamiltonian is not simply the opposite sign of the fourth-order
Lagrangian and the additional contributions in the interaction
Hamiltonian also do not have the symmetry between the 
adiabatic mode
and entropy mode. Thus the momentum dependence of the trispectrum
becomes different for adiabatic, entropy and mixed components.

Therefore the third order is a special case where the shape of the bispectrum
is not changed by the entropy modes. This is because, at third order, the
interaction Hamiltonian only arises by the boost which preserves the symmetry
between the adiabatic and the entropy modes.
At higher orders, we expect that the momentum dependence
of the $n$-point function induced by the entropy modes is different from
the pure adiabatic contribution. This would be crucial
to distinguish between single field and multi-field models
by the shape dependence of the $n$-point functions.

\section{\label{sec:CONCLUSIONS}
Conclusions}

In the multi-field DBI-inflation model,
it had been shown that the sound
speeds for the adiabatic and entropy perturbations
are the same
\cite{Langlois:2008wt}.
It was also shown that the momentum dependence of the three-point
function of the final curvature perturbation remains the same as in the single-field case, that is,
the components of the three-point function including the
entropy perturbations only change the amplitude
of the three-point function from the
purely adiabatic component.
This is because there exists a symmetry under the
exchange between the adiabatic and entropy modes in
the second- and third-order actions for $Q_n$ at leading
order in slow-roll and in the small sound speed limit.

In this paper, as a natural extension of these works,
we studied the non-Gaussianity from the contact interaction trispectrum
in the multi-field DBI-inflation model.
We first derived the fourth-order action
for the perturbations based on the usual ADM formalism
(Eq.(\ref{two_field_lead_dbi_four})).
It is written in terms of the adiabatic and entropy
perturbations in the small sound speed limit
and at leading order in the slow-roll expansion.

After deriving the relation between
the interaction Lagrangian and the interaction
Hamiltonian, which can be applied to a fairly general
two-field model, we obtained the fourth-order
interaction Hamiltonian
(Eq.~(\ref{two_field_lead_dbi_hamil_four})).
It is worth noting that while the cubic part of
the interaction Hamiltonian is the opposite sign
of the Lagrangian density, this is generally not true
at higher orders \cite{Huang:2006eh}.

Using this interaction Hamiltonian, we derived the connected four-point
function coming from the intrinsic fourth-order contact interaction, in the small sound speed limit and at leading order in
the slow-roll expansion.
In these approximations, in addition to the purely
adiabatic four-point function
$\langle Q_\sigma(\mathbf{k_1})Q_\sigma(\mathbf{k_2})
Q_\sigma(\mathbf{k_3})Q_\sigma(\mathbf{k_4})
\rangle$, there exists a purely entropic component
$\langle Q_s(\mathbf{k_1})Q_s(\mathbf{k_2})
Q_s(\mathbf{k_3})Q_s(\mathbf{k_4})
\rangle$ and a mixed component
$\langle Q_\sigma (\mathbf{k_1})Q_\sigma (\mathbf{k_2})
Q_s(\mathbf{k_3})Q_s(\mathbf{k_4})
\rangle$. It was shown that the purely entropic
and the mixed components have different momentum dependence
from the purely adiabatic component
(Eqs.~(\ref{4point_pureadiabatic}),
(\ref{4point_pureentropy}) and (\ref{4point_mixed}) ).
Because of this it was shown that the momentum
dependence of the four-point function
of the comoving curvature perturbations
is affected by the entropy modes
(Eq.~(\ref{4point_curv})), and the shape is different from the
single-field case.
In contrast to the shape of the bispectrum, which does not
distinguish the multi-field DBI-inflation model from the single-field
DBI-inflation model, the CMB
trispectrum can provide a useful discriminator for
the multi-field DBI-inflation model.

We also derived the fourth-order action
for the perturbations by an alternative and simpler
method. Since the DBI action describes the motion
of a brane in a higher dimensional spacetime,
the action for the perturbations can be obtained
by a Lorentz boost from the frame where the brane
is at rest in the background
to the frame where the brane moves at the
velocity $v$ with $v^2 = 2 f_0 X_0 = f_0 \dot{\sigma}^2_0$.
In the small sound speed limit
($ v \to 1$), the actions up to fourth order
in the frame where the brane is moving were calculated
(Eqs.~(\ref{2ndorderaction_boost}),
(\ref{3rdorderaction_boost}) and (\ref{4thorderaction_boost}))
and it was shown that they coincide with the ones obtained
by the usual method (Eqs.~(\ref{two_field_lead_dbi_sec}),
(\ref{two_field_lead_dbi_thir}) and
(\ref{two_field_lead_dbi_four})).
From this derivation, we found that, at third order,
the interaction Hamiltonian arises purely by the boost
and it has the symmetry under the exchange between the
adiabatic mode and entropy mode. This is because the
boost does not distinguish between them. At fourth
order, there exists the intrinsic fourth-order action
in the rest frame of the brane. This breaks the symmetry.
In addition, the interaction Hamiltonian is not just the
opposite sign of the Lagrangian. The additional terms in
the interaction Hamiltonian also break the symmetry.
Hence the trispectrum coming from the entropy modes has a different
shape dependence from the one coming from the adiabatic modes.

In order to calculate the effect of
the entropy perturbation on the curvature perturbation, we need
to specify a model that describes how the entropy
perturbation is converted to the curvature perturbation.
In this paper, we modeled this transfer by a
transfer function $T_\mathcal{RS}$.
It would be interesting to study this mixing in specific
string theory motivated models.

In this paper, we considered the trispectrum coming from the intrinsic
fourth-order contact interaction. However, as it was shown recently in
\cite{Gao:2009gd,Chen:2009bc,Us}, there are other important
contributions for the trispectrum coming from the interactions at a
distance such as the exchange of scalar particles.
In particular,
Ref. \cite{Gao:2009gd} calculated a part of the trispectrum coming from the entropy modes, i.e. $\langle\mathcal{R}(\mathbf{k_1})\mathcal{R}(\mathbf{k_2})
\mathcal{R}(\mathbf{k_3}) \mathcal{R}(\mathbf{k_4})\rangle\propto T_\mathcal{RS}^4\langle Q_s(\mathbf{k_1})Q_s(\mathbf{k_2})
Q_s(\mathbf{k_3})Q_s(\mathbf{k_4})\rangle$, when an adiabatic scalar
particle is exchanged. We will present the full leading order trispectrum in the multi-field DBI-inflation model in a separate publication \cite{Us3}.

\begin{acknowledgments}
SM and FA are supported by the Japanese Society for the Promotion of
 Science (JSPS). KK is supported by ERC, RCUK and STFC.
TT is supported by Grants-in-Aid for Scientific Research, Nos. 19540285,
 17340075 and 21244033. SM is grateful to the ICG, Portsmouth for their hospitality
when part of this work was done. The authors also would like to mention that
discussions during the GCOE/YITP workshop YITP-W-09-01 on "Non-linear
 cosmological perturbations" were useful to complete this work.
\end{acknowledgments}

\end{document}